\pdfoutput=1

\documentclass[aip,rsi,reprint,graphicx]{revtex4-1}
\usepackage{amsmath}
\usepackage{amssymb}
\usepackage[dvipsnames]{xcolor}
\usepackage{graphicx}
\usepackage[toc,page]{appendix}
\usepackage{dcolumn}
\usepackage{bm}
\usepackage{float}

\begin{document}

\preprint{}

\title{Laser frequency stabilization using a commercial wavelength meter}

\author{Luc Couturier}

\affiliation{Hefei National Laboratory for Physical Sciences at the Microscale and Shanghai Branch, University of Science and Technology of China, Shanghai 201315, China}

\affiliation{CAS Center for Excellence and Synergetic Innovation Center in Quantum Information and Quantum Physics, University of Science and Technology of China, Shanghai 201315, China}

\author{Ingo Nosske}

\affiliation{Hefei National Laboratory for Physical Sciences at the Microscale and Shanghai Branch, University of Science and Technology of China, Shanghai 201315, China}

\affiliation{CAS Center for Excellence and Synergetic Innovation Center in Quantum Information and Quantum Physics, University of Science and Technology of China, Shanghai 201315, China}

\author{Fachao Hu}

\affiliation{Hefei National Laboratory for Physical Sciences at the Microscale and Shanghai Branch, University of Science and Technology of China, Shanghai 201315, China}

\affiliation{CAS Center for Excellence and Synergetic Innovation Center in Quantum Information and Quantum Physics, University of Science and Technology of China, Shanghai 201315, China}

\author{Canzhu Tan}

\affiliation{Hefei National Laboratory for Physical Sciences at the Microscale and Shanghai Branch, University of Science and Technology of China, Shanghai 201315, China}

\affiliation{CAS Center for Excellence and Synergetic Innovation Center in Quantum Information and Quantum Physics, University of Science and Technology of China, Shanghai 201315, China}

\author{Chang Qiao}

\affiliation{Hefei National Laboratory for Physical Sciences at the Microscale and Shanghai Branch, University of Science and Technology of China, Shanghai 201315, China}

\affiliation{CAS Center for Excellence and Synergetic Innovation Center in Quantum Information and Quantum Physics, University of Science and Technology of China, Shanghai 201315, China}

\author{Y. H. Jiang}

\email{jiangyh@sari.ac.cn}

\affiliation{Shanghai Advanced Research Institute, Chinese Academy of Sciences, Shanghai 201210, China}

\author{Peng Chen}

\email{peng07@ustc.edu.cn}

\affiliation{Hefei National Laboratory for Physical Sciences at the Microscale and Shanghai Branch, University of Science and Technology of China, Shanghai 201315, China}

\affiliation{CAS Center for Excellence and Synergetic Innovation Center in Quantum Information and Quantum Physics, University of Science and Technology of China, Shanghai 201315, China}

\author{Matthias Weidem\"uller}

\email{weidemueller@uni-heidelberg.de}

\affiliation{Hefei National Laboratory for Physical Sciences at the Microscale and Shanghai Branch, University of Science and Technology of China, Shanghai 201315, China}

\affiliation{CAS Center for Excellence and Synergetic Innovation Center in Quantum Information and Quantum Physics, University of Science and Technology of China, Shanghai 201315, China}

\affiliation{Physikalisches Institut, Universit\"at Heidelberg, Im Neuenheimer Feld 226, 69120 Heidelberg, Germany}

 \date{\today}

\begin{abstract}
We present the characterization of a  laser frequency stabilization scheme using a state-of-the-art wavelength meter based on solid Fizeau interferometers. For a frequency-doubled Ti-sapphire laser operated at 461 nm, an absolute Allan deviation below $10^{-9}$ with a standard deviation of $1$ MHz over 10\,h is achieved. Using this laser for cooling and trapping of strontium atoms, the wavemeter scheme provides excellent stability in single-channel operation. Multi-channel operation with a multimode fiber switch results in fluctuations of the atomic fluorescence correlated to residual frequency excursions of the laser. The wavemeter-based frequency stabilization scheme can be applied to a wide range of atoms and molecules for laser spectroscopy, cooling and trapping.
\end{abstract}
\maketitle


\section{Introduction}

Frequency stabilization of a laser source is indispensable for a variety of applications in laser spectroscopy\cite{demtroder2008laser}, quantum optics\cite{scully1999} and high-accuracy metrology\cite{ludlow2015optical}. In particular, it is essential for laser cooling and trapping of atoms, ions and molecules, where the Doppler cooling effect is essential\cite{metcalf2007}. A great number of spectroscopic methods\cite{Su2014,Zi2017,Harris2006,Bjorklund1983,McCarron2008} have been developed over the years to stabilize the laser frequency to atomic or molecular lines. However, the convenience and availability of vapor cells for some elements or molecules are limited. As an example, for elements with high melting points like Li, Sr, Yb and Er, heat pipes or atomic beams have to be used\cite{Li2004,Bridge2009}. Although hollow-cathode lamps are compact alternatives for such cases, they usually suffer short lifetimes and large pressure broadening\cite{Norcia2016,Lee2014}. Spectra of molecules such as iodine are also used even though the transition strengths are rather weak\cite{Talvitie1998,Vasilyev2011}.

A different approach which is not wavelength specific is to use a reference cavity\cite{tonyushkin2007,Rossi2002a,Uetake2009} which exhibits a series of optical resonant peaks with a typical free spectral range (FSR) in the GHz range. Each resonant peak can serve as a reference for frequency stabilization using, for instance, a Pound-Drever-Hall technique\cite{Black2001}. However, locking to a reference cavity may result in noticeable frequency drifts depending on the mechanical design and the thermal expansion coefficients of the optical resonators. The frequency drift can vary from hundreds of MHz/hour to $<10\,$kHz/day\cite{alnis2008}. To eliminate the frequency drift, the cavity length is often further stabilized to an external spectral line or carefully isolated, at the expense of more complexity.

Another alternative, which is not suffering from a finite frequency capture range as the methods discussed before, is to use a high-accuracy wavelength meter (WLM)\cite{kobtsev2007,Metzger2017}. It requires only a little amount of light power, typically a few tens of $\mu$W, and can be used over a broad spectrum, typically few hundreds of nanometers. WLMs based on solid Fizeau interferometers offer a specified accuracy down to the MHz level in a broad spectral range\cite{kobtsev2007,saleh2015}. Such a level of accuracy is already of interest for laser cooling atoms with broad lines, such as Yb, Sr and Hg\cite{Sansonetti2005}. So far, there have been few applications of WLMs as frequency stabilization schemes\cite{kobtsev2007,saleh2015,saakyan2015,Barwood2012}. Previous work showed that a frequency instability of order $10^{-10}$ for a single laser can be achieved by measuring the Allan deviation of the frequency of a near-infrared fiber laser stabilized by a high-accuracy WLM with a single-mode fiber\cite{saleh2015}. We extend these studies to applications in atomic cooling and trapping, and we reveal limitations in multi-laser operation using multi-mode fiber switches.

We present a comprehensive characterization of the frequency stabilization scheme with a commercial high-accuracy WLM (HighFinesse WSU-10). We quantify the frequency stability by the Allan deviation\cite{riley2008handbook} over time intervals between 0.1 and 1000~s, using a laser stabilized to a narrow atomic resonance as a reference. The performance of the frequency stabilization scheme is analyzed using a Ti-sapphire laser and we find that the uncertainty of the provided frequency readout is the major limit of the frequency stabilization scheme. In order to demonstrate its application to an atom cooling experiment, we apply this stabilization scheme to the laser used for a magneto-optical trap (MOT) of strontium atoms and analyze its performance, comparing single- and multi-laser operation schemes.
   
\section{Setup}
\subsection{Fizeau wavelength meter}

Compared to other instruments for wavelength determination\cite{Scholl2004,Banerjee2001}, such as grating spectrometers or Michelson interferometers, WLMs based on solid Fizeau interferometers offer high accuracy and reliability\cite{reiser1988,gray1986,Gardner1985,Kajava1993,Rogers1982,Kajava1994}. 

The optical unit of the WLM used in our laboratory (HighFinesse WSU-10) comprises two sets of multiple interferometers (in total 6 Fizeau interferometers) with a dedicated mechanical design and two CCD line arrays. The optical unit also includes two fiber coupling inputs, one for wavelength determination and the other for calibration. The measured wavelength is calculated by comparison of the recorded interference pattern with the one of the calibration laser, which is a frequency stabilized HeNe laser with a maximum stability of $2\times 10^{-8}$ in our case. As the photodiode arrays are very sensitive, input laser powers of a few tens of $\mu$W with an exposure time of $1\,$ms are sufficient for operation.

The WLM supports multi-channel frequency measurements of up to eight lasers simultaneously, when equipped with an external multi-mode fiber switch. It can provide electronic feedback loops for two lasers via built-in proportional-integral-derivative (PID) modules. When using a single-mode fiber (single-laser operation), the specified absolute accuracy is $\pm10\,$MHz (within $\pm 200\,$nm around the calibration wavelength) with the specified measurement resolution being $1\,$MHz over the standard spectral range ($350-1120\,$nm). The latter frequency precision is wavelength dependent. In multi-laser operation (multi-mode), the specified accuracy is specified as $\pm 100\,$MHz. 

\begin{figure*}
\includegraphics[width=0.9\textwidth]{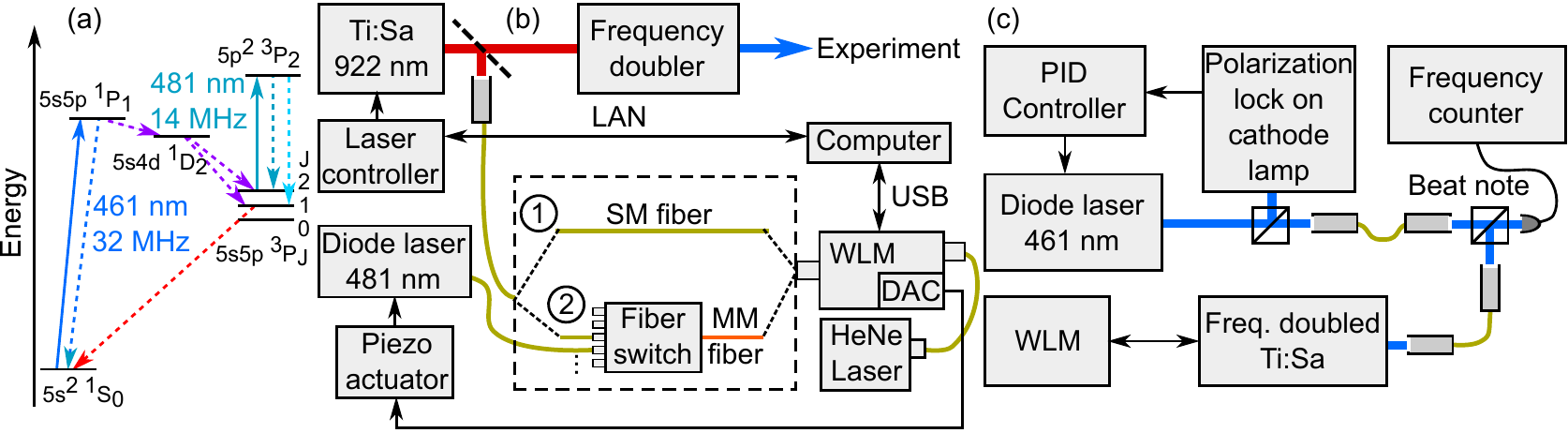}
\caption{Schematic of the experimental setup for laser stabilization using the Fizeau wavelength meter (WLM) and relevant energy levels for laser cooling of strontium atoms. (a)  Relevant level structure of strontium for laser cooling. The $5s^2\,^1\mathrm{S}_0\rightarrow 5s5p\,^1\mathrm{P}_1$ transition is used for laser cooling and the $5s5p\,^3\mathrm{P}_2\rightarrow 5p^2\,^3\mathrm{P}_2$ is used to repump atoms accumulated in the $5s5p\,^3\mathrm{P}_2$ metastable state. (b) Laser stabilization scheme of  Ti-sapphire laser ($922\,$nm) and a diode laser ($481\,$nm). A fraction of the light is sent to the WLM using \textcircled{\small{1}} a single-mode optical fiber (yellow) or \textcircled{\small{2}} a multi-mode fiber switch and a short multi-mode fiber (orange).  The calculated wavelength signal of the Ti-sapphire laser is sent from the the computer over the local-area network to the laser controller of the Ti-sapphire laser.  The diode laser is stabilized using the integrated digital-to-analog converter of the WLM which drives the piezo actuator of the diode laser. A HeNe laser calibrates the WLM through another fiber coupling port. (c) Beatnote setup for frequency uncertainty characterization of the stabilized Ti-sapphire laser.}
\label{ExpSetup}
\end{figure*}

\subsection{Laser system for strontium cooling and trapping}

The main purpose of our experiment is to optically cool and trap atomic strontium\cite{nosske2017}. The relevant level scheme is shown in Fig.~\ref{ExpSetup} (a). A Ti-sapphire laser (M2 lasers SolsTiS-PSX-1600-F) with a wavelength of $922\,$nm is frequency doubled (M2 lasers ECD-X) to generate the cooling laser light at $461\,$nm for strontium atoms. This cooling laser is near-resonant to the transition $5s^2\,^1\mathrm{S}_0\rightarrow 5s5p\,^1\mathrm{P}_1$ with a natural width of $32\,$MHz. Since the cooling transition is not closed, population in the excited state $^1\mathrm{P}_1$ slowly leaks to the long-lived metastable state $^3\mathrm{P}_2$ which creates loss of atoms in the laser cooling process. To pump atomic population back to the ground state, we employ a diode laser (Toptica DL100) at a wavelength of $481\,$nm to address the transition $5s5p\,^3\mathrm{P}_2\rightarrow 5p^2\,^3\mathrm{P}_2$. 

The Ti-sapphire laser and diode laser (in the following called cooling and repumping laser, respectively) can be stabilized independently or simultaneously by the WLM, as shown in Fig.~\ref{ExpSetup} (b). A small portion of light power of the Ti-sapphire laser is coupled to the WLM either via a single-mode fiber (\textcircled{\small{1}}, yellow) or a multi-mode channel (\textcircled{\small{2}}). The latter consists of a multi-channel fiber switch (multi-mode) and a short piece of multi-mode fiber ($20\,$cm, orange) which accepts a broad wavelength range.  We find that the use of a rather short multi-mode fiber can effectively reduce the uncertainty of the wavelength measurement by reducing additional phase shifts induced by temperature fluctuations of the fiber. The WLM and switch are enclosed in a box to mitigate temperature fluctuations. 

The channels are read sequentially and the total acquisition rate of the WLM is less than $1\,$kHz, thus the calculation of the wavelength is performed with the proprietary software on a computer which communicates with the WLM by a USB connection. The software can also generate electronic feedback signals for laser frequency stabilization. The Ti-sapphire laser controller contains a frequency stabilization function. Therefore, the digital wavelength signal can be directly sent from the controller via the network. Frequency stabilization of the repumping laser is implemented by a digital-analog-converter (DAC,  WLM built-in module) which converts the digital frequency signal to an analog voltage signal which is used to control the diode laser's frequency.

\section{Characterization of frequency stability}
\label{WLMreading}

\subsection{Readout stability of the wavelength meter}
\label{readoutstability}

First, we characterize the frequency readout uncertainty of the WLM over time. To do so, we measure the frequency of an ultra-stable diode laser (Toptica DLC TA Pro  at $689\,$nm). This diode laser, not shown in Fig.~\ref{ExpSetup}, is frequency stabilized to an ultra-low expansion (ULE) passive cavity (Stable Laser Systems) with a finesse around $200\,000$ by a PDH lock. The resulting linewidth of the laser is less than $1\,$kHz over 1\,s, and the long-term drift of the ULE passive cavity is below $10\,$kHz per day. Thus, the frequency stability of this laser is much better than the specified resolution of the WLM, and frequency noise of the laser can be neglected\cite{saleh2015}.

\begin{figure}
\centering 
\includegraphics[width=0.95\linewidth]{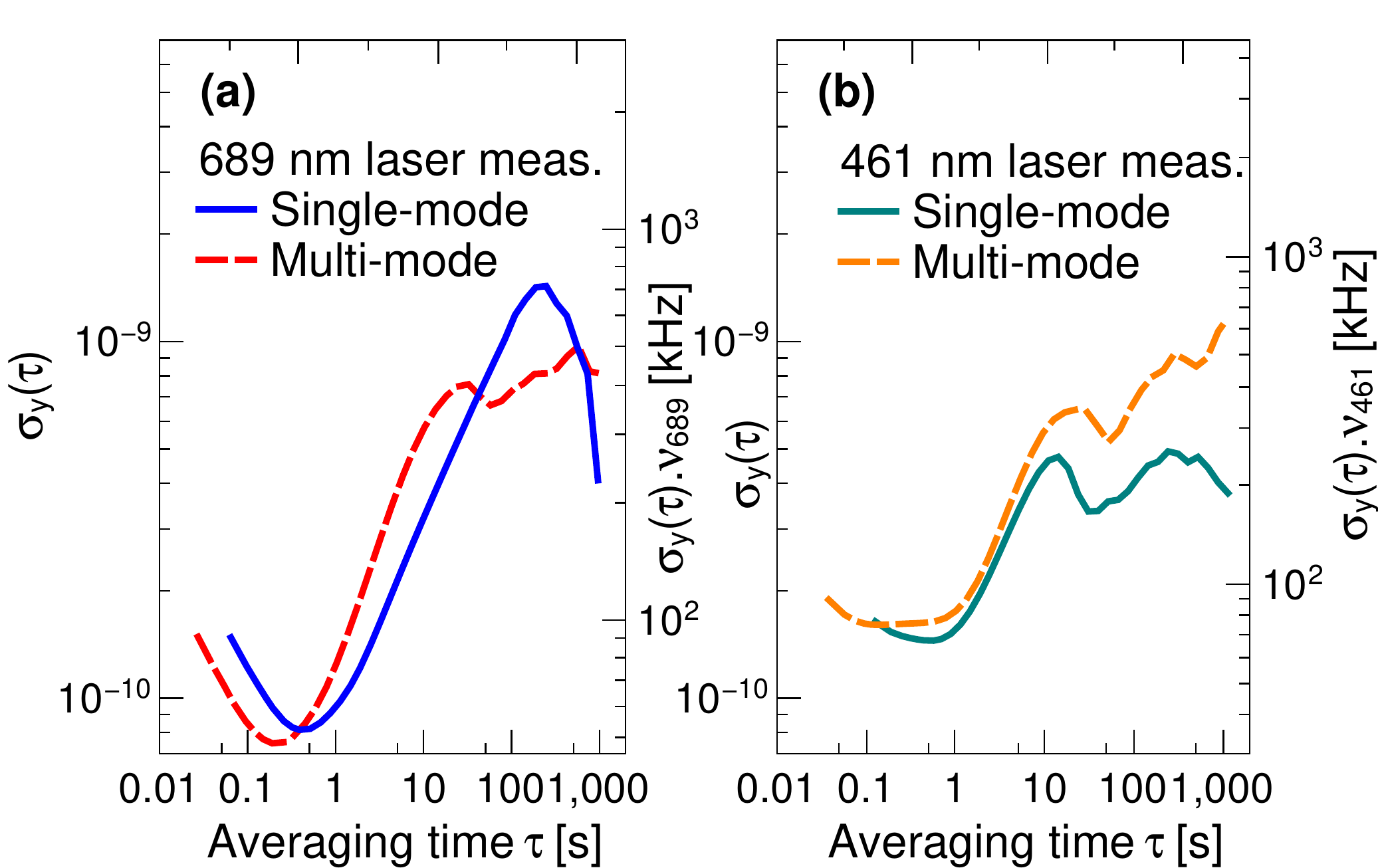}
\caption{Stability characterization of the wavelength meter (WLM) frequency read-out. (a) Allan deviation of frequency measurement of an ultra-stable $689\,$nm diode laser. (b) Allan deviation of the frequency measurement of a $461\,$nm diode laser stabilized to the strontium atomic line. The read-out stability of the WLM via a single-mode fiber (single-channel) and a multi-mode switch (multi-channel) are compared. $\nu_{689}$ and $\nu_{461}$ denote the central frequencies of ultra-stable diode laser and a $461\,$nm diode laser, respectively. Each measurement is taken over 10 hours.
}
\label{AllanWLM}
\end{figure}

As shown in Fig.~\ref{AllanWLM} (a), the Allan deviation $\sigma_y(\tau)$ decreases due to averaging effects down to a minimum of $8\times 10^{-11}$ for an averaging time of $\sim300\,$ms. The Allan deviation then increases up to a maximum of $1.4\times 10^{-9}$ around $200\,$s. The decrease at even longer time scales is due to averaging effects of the frequency as the reading does not present systematic drift. A recalibration of the WLM is done every $30\,$min and helps to further improve the stability, different to previous work\cite{saleh2015}. The standard deviation of the frequency distribution over $10\,$h is about $500\,$kHz. In multi-channel operation, the Allan deviation has a maximum of $1\times 10^{-9}$ around $600\,$s and does not tend to decrease further. The slope of the curve in the double-logarithmic representation for times up to a few hundred seconds is consistent with a random walk behavior without indication for a continuous drift. The frequency distribution yields a $1.2\,$MHz standard deviation over a $10\,$h measurement time. We have compared our findings to similar data provided by the manufacturer HighFinesse\cite{privcomm} with an improved version of the WLm, which are consistent with the data that we present here.

\subsection{Characterization of wavemeter-based frequency stabilization scheme}
\label{part:stabCharac}

To analyze the frequency uncertainty of the Ti-sapphire laser at $461\,$nm stabilized to the WLM, we have set up a measurement of the beatnote with a diode laser stabilized to a strontium atomic resonance at the same wavelength, as depicted in Fig.~\ref{ExpSetup} (c). The diode laser (Moglabs CEL002) is frequency stabilized to the strontium transition $5s^2\,^1\mathrm{S}_0\rightarrow 5s5p\,^1\mathrm{P}_1$, using polarization spectroscopy with a hollow cathode lamp, based on an existing work\cite{shimada2013}. This diode laser has a linewidth of about $500$ kHz over a time interval of $1\,$s, measured with an optical spectrum analyzer (Sirah EagleEye). The Allan deviation of the laser frequency determined similarly as described in the previous section is shown on Fig.~\ref{AllanWLM} (b). The Allan deviation remains within a factor of 3 for all timescale compared to the Allan deviation of the ultrastable laser. Over the 10\,h measurement, the standard deviation is $700\,$kHz and $3.7\,$MHz for single- and multi-channel operation, respectively. The stability measurement is thus limited by the stability of the WLM. Therefore the laser can be employed as a reference for the beat note measurement with another laser that is stabilized by the WLM. An accurate frequency counter is used to measure the frequency difference between the stabilized Ti-sapphire laser and the diode laser resulting from the beat signal (at a frequency offset around 160\,MHz).

\begin{figure}
\includegraphics[width=0.95\linewidth]{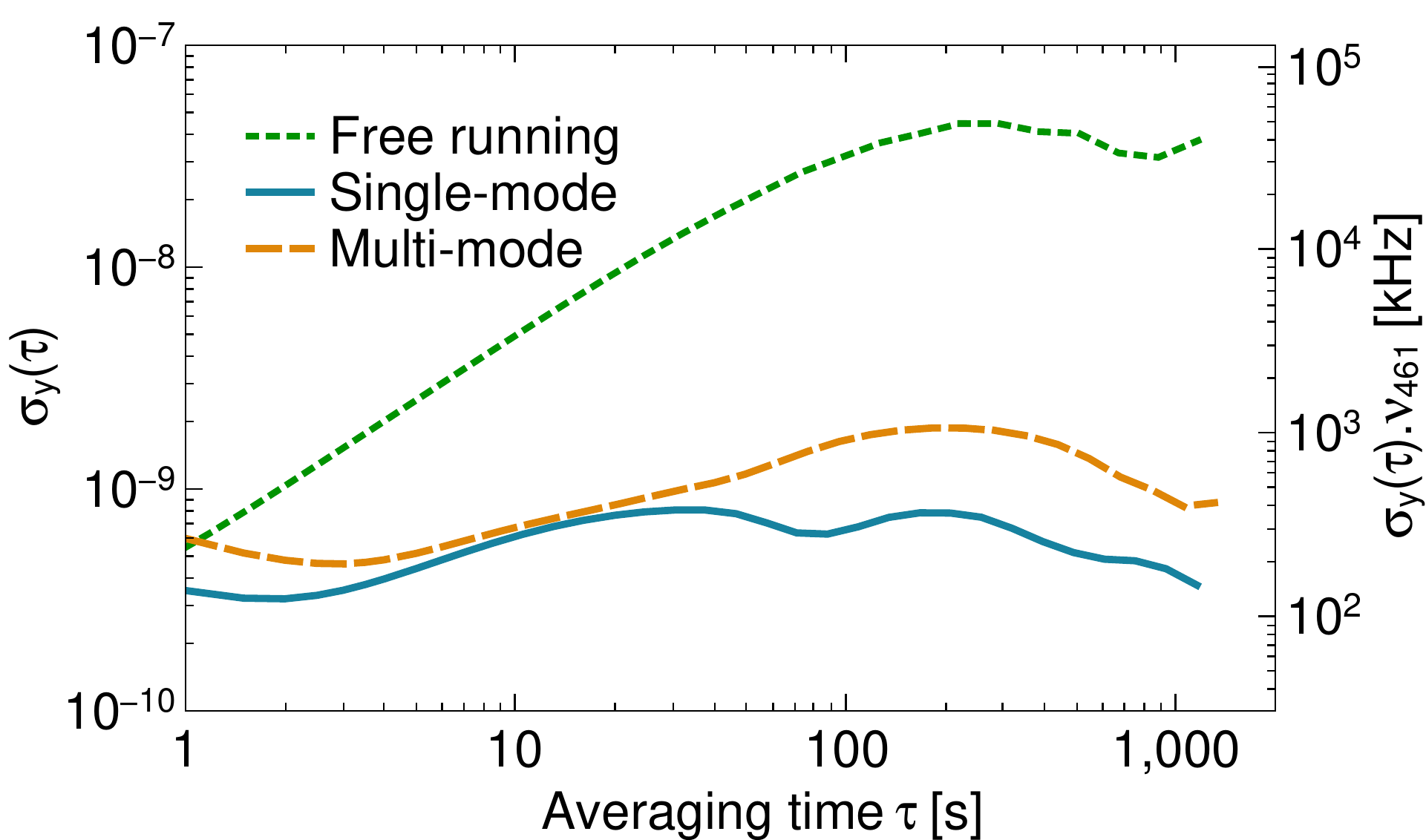}
\caption{ Frequency stability characterization of the frequency-doubled Ti-sapphire laser stabilized to the WLM. Allan deviations are calculated from the beatnote between the frequency-doubled Ti-sapphire laser and the 461 nm diode laser stabilized to the strontium atomic line with polarization spectroscopy. The Ti-sapphire laser is stabilized to the WLM via a single-mode fiber, a multi-mode switch or is free running. The measurement rate of the beatnote is $2\,$Hz. Each measurement is taken over 10 hours.}
 \label{AllanMOGLL}
\end{figure}

We compare two scenarios: the Ti-sapphire is stabilized and coupled to the WLM with a single-mode fiber (single-channel), and, alternatively, with a multi-mode fiber(multi-channel) which allows one to stabilize multiple lasers simultaneously. The Allan deviation $\sigma_y(\tau)$ calculated from the beatnote measurements is plotted in Fig.~\ref{AllanMOGLL}, where the Allan deviation for the free-running of frequency-doubled Ti-sapphire laser is also shown. The free running laser shows a linear drift of $53\,$MHz/hour. When operated with the WLM frequency stabilization in single-channel operation, this drift is suppressed resulting to an Allan deviation staying below $8\times 10^{-10}$ over time intervals between 1\,s and 1000\,s. The steady increase of the Allan deviation in multi-channel operation might indicate a small residual drift, yet the Allan deviation does not exceed $2\times 10^{-9}$. In fact, the resulting Allan deviation in both operation modes is comparable to the genuine stability of the WLM itself (see Fig.~\ref{AllanWLM}), suggesting that the achievable frequency stability is actually limited by the WLM readout. The standard deviation of the frequency distribution is $1.1\,$MHz in single-channel operation and $1.8\,$MHz in multi-channel operation, respectively, over a period of 10\,h.

\section{Laser cooling and trapping of strontium}
\label{sec:MOT}

These measured frequency deviations are much smaller than the width of the strontium transition ($32\,$MHz). Thus, the WLM stabilization scheme provides favorable conditions for employing the laser system in a laser cooling and trapping experiment. Details of our Sr-MOT, which combines a 2D-MOT with a 3D-MOT, are described in Ref. \cite{nosske2017}. In single-laser operation, only the Ti-sapphire laser is frequency stabilized to the WLM via a single-mode fiber, and no repumping laser is used which results in a MOT containing $10^{6}$ atoms. In multi-laser operation, we stabilize both the cooling and the repumper laser to the WLM using the multimode-fiber switch (see Fig. \ref{ExpSetup}). Under these conditions, about $10^{7}$ atoms are being trapped. We simultaneously record the frequency deviations of the cooling laser, as described in the previous section, and the fluorescence of the trapped atoms using a photodiode over a period of 5000\,s. There is no simple relation between the fluorescence intensity and the atom number. The frequency of the repumping laser is not monitored as the repumping line is power-broadened to roughly 100\,MHz\cite{Sansonetti2010}, resulting in a negligible influence of the repumping laser frequency stability on the MOT performance.

\begin{figure}
\centering
\includegraphics[width=0.95\linewidth]{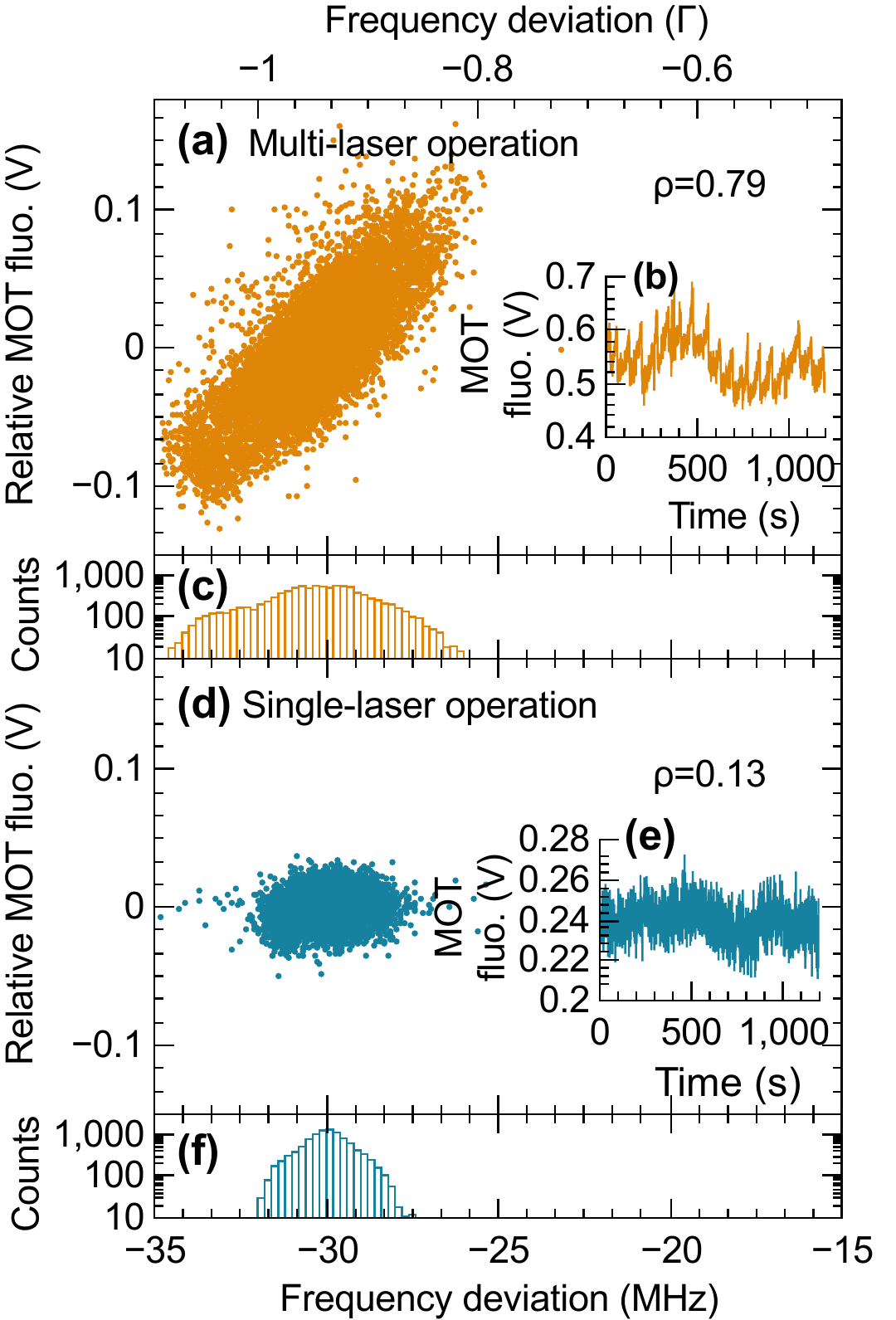}
\caption{Fluorescence of the strontium MOT and its correlation to frequency deviation of the cooling laser stabilized by the wavelength meter (WLM).  (a) and (d): Correlations of the fluorescence intensity with the frequency of the cooling laser in the multi-laser and single-laser operation, respectively, measured over $5000\,$s. The frequency is referenced to the atomic transition, and the fluorescence signal is relative to the its mean value.  (c) and (f) are the histograms of the frequency distributions of (a) and (d). (b) and (e) are excerpts of the total fluorescence trace over time.}
\label{Fig4:correlations}
\end{figure}

Fig.~\ref{Fig4:correlations} (a) shows the fluorescence intensity of the MOT as a function of the frequency deviation of the cooling laser in multi-laser operation. There is a clear positive correlation with a correlation coefficient\cite{Barlow1989} of $\rho=0.79$. In the histogram of Fig.~\ref{Fig4:correlations} (c), the frequency distribution fits to a Gaussian distribution with $\sigma=1.6\,$MHz, with an additional shoulder at $-33\,$MHz. This width agrees with the standard deviation determined for the laser stability in Sec.~\ref{part:stabCharac}.

An excerpt of the fluorescence intensity fluctuations over time is given in Fig.~\ref{Fig4:correlations} (b). Two periodic structures are observed. The oscillation with a period of $\sim20\,$min is correlated to ambient temperature fluctuations, and might thus indicate a slight temperature dependence of the WLM readout in multi-channel operation. The spikes which appear every 60\,s are related to a periodic perturbation in the WLM readout, which can also directly been seen in its frequency monitor. This regular feature might be the cause for the increase of the WLM's Allan deviation beyond 50\,s as shown in Fig.~\ref{AllanMOGLL}.

Fig.~\ref{Fig4:correlations} (d) shows the fluorescence intensity of the MOT as a function of the frequency deviation of the cooling laser in the single-channel operation. Under these conditions, the correlation coefficient is $\rho=0.13$ which suggests that the fluorescence intensity of the MOT is uncorrelated with the frequency deviation of the cooling laser. The frequency distribution exhibits a standard deviation $\sigma=0.8\,$MHz, as depicted in the histogram in Fig.~\ref{Fig4:correlations} (f), again agreeing well with the genuine laser deviation as presented in Sec.~\ref{part:stabCharac}. An excerpt of the fluorescence intensity fluctuation over time (see Fig.~\ref{Fig4:correlations} (b)). The observed fluctuations of the fluorescence amounts to roughly 5\%, well above the electronic noise of the detector and on the order of the laser intensity fluctuations. There is no indication of periodic behavior as previously seen in multi-channel operation.

\section{Conclusion}

We have characterized a laser frequency stabilization scheme using a commercial WLM.  In single-channel operation, the frequency standard deviation of the WLM reading is less than  $1\,$MHz at $689\,$nm over $10$ hours. The frequency stability of a frequency-doubled Ti-sapphire at 461\,nm locked to the WLM in single-channel mode shows similar performance with a frequency standard deviation of $1.1\,$MHz over $10\,$h and a short term stability below $10^{-9}$. In multi-channel operation, the standard deviation of the laser frequency yields $1.8\,$MHz with a short-term stability of $1\times 10^{-8}$. In comparison to previous work\cite{saleh2015,saakyan2015}, no continuous drift of the WLM frequency reading is observed due to autocalibration of the WLM. The WLM frequency stabilization scheme is used to realize a MOT of strontium atoms. In multi-laser operation, we observe a direct correlation between frequency and fluorescence fluctuations. In single-laser operation, the fluorescence intensity fluctuations are nearly uncorrelated with the frequency uncertainty of the cooling laser. Recently, broadband single-mode fiber switches based on photonic crystal fibers became available for commercial WLMs\cite{HighFinesse}. They can preserve single-mode operation over a broader spectrum than single-mode fiber switches. We anticipate that with these switches the WLM stabilization scheme will perform as well as in single-laser operation.

Frequency stabilization using WLMs features great advantages as compared to other schemes based on, e.g., atomic spectroscopy or cavity resonances. Thanks to the unlimited frequency capture range, combined with a sufficient frequency stability, we can operate the Sr MOT for an entire day in single-laser and multi-laser stabilization mode without a single loss of trapped atoms and with exquisite long-term constancy of the number of trapped atoms \cite{nosske2017}. In a recent extension of our setup, we have integrated a third laser for Rydberg excitation (frequency-doubled dye laser at 318\,nm, Sirah Matisse 2 DX) into the frequency stabilization scheme allowing for accurate determination of Rydberg line frequencies without compromising MOT performance. Based on this experience, we expect a broader range of applications of WLM-based frequency stabilization schemes in atomic and molecular laser spectroscopy, cooling and trapping.

\begin{acknowledgments}
We thank Jan Blume for support at the early stage of the experiment, and Florian Karlewski from HighFinesse for providing us with an Allan deviation measurement and helpful remarks. M.W.'s research activities in China are supported by the 1000-Talent-Program of the Chinese Government. The work was supported by the National Natural Science Foundation of China (Grant Nos. 11574290 and 11604324). Y.H.J. also acknowledges support under Grant Nos. 11420101003 and 91636105.
\end{acknowledgments}

\bibliography{manuscript_WL_V13}

\begin{thebibliography}{41}%
\makeatletter
\providecommand \@ifxundefined [1]{%
 \@ifx{#1\undefined}
}%
\providecommand \@ifnum [1]{%
 \ifnum #1\expandafter \@firstoftwo
 \else \expandafter \@secondoftwo
 \fi
}%
\providecommand \@ifx [1]{%
 \ifx #1\expandafter \@firstoftwo
 \else \expandafter \@secondoftwo
 \fi
}%
\providecommand \natexlab [1]{#1}%
\providecommand \enquote  [1]{``#1''}%
\providecommand \bibnamefont  [1]{#1}%
\providecommand \bibfnamefont [1]{#1}%
\providecommand \citenamefont [1]{#1}%
\providecommand \href@noop [0]{\@secondoftwo}%
\providecommand \href [0]{\begingroup \@sanitize@url \@href}%
\providecommand \@href[1]{\@@startlink{#1}\@@href}%
\providecommand \@@href[1]{\endgroup#1\@@endlink}%
\providecommand \@sanitize@url [0]{\catcode `\\12\catcode `\$12\catcode
  `\&12\catcode `\#12\catcode `\^12\catcode `\_12\catcode `\%12\relax}%
\providecommand \@@startlink[1]{}%
\providecommand \@@endlink[0]{}%
\providecommand \url  [0]{\begingroup\@sanitize@url \@url }%
\providecommand \@url [1]{\endgroup\@href {#1}{\urlprefix }}%
\providecommand \urlprefix  [0]{URL }%
\providecommand \Eprint [0]{\href }%
\providecommand \doibase [0]{http://dx.doi.org/}%
\providecommand \selectlanguage [0]{\@gobble}%
\providecommand \bibinfo  [0]{\@secondoftwo}%
\providecommand \bibfield  [0]{\@secondoftwo}%
\providecommand \translation [1]{[#1]}%
\providecommand \BibitemOpen [0]{}%
\providecommand \bibitemStop [0]{}%
\providecommand \bibitemNoStop [0]{.\EOS\space}%
\providecommand \EOS [0]{\spacefactor3000\relax}%
\providecommand \BibitemShut  [1]{\csname bibitem#1\endcsname}%
\let\auto@bib@innerbib\@empty
\bibitem [{\citenamefont {Demtr{\"o}der}(2008)}]{demtroder2008laser}%
  \BibitemOpen
  \bibfield  {author} {\bibinfo {author} {\bibfnamefont {W.}~\bibnamefont
  {Demtr{\"o}der}},\ }\href@noop {} {\emph {\bibinfo {title} {Laser
  Spectroscopy Volume 1 Basic Principles}}},\ Vol.~\bibinfo {volume} {1}\
  (\bibinfo  {publisher} {Springer},\ \bibinfo {year} {2008})\BibitemShut
  {NoStop}%
\bibitem [{\citenamefont {Scully}\ and\ \citenamefont
  {Zubairy}(1999)}]{scully1999}%
  \BibitemOpen
  \bibfield  {author} {\bibinfo {author} {\bibfnamefont {M.~O.}\ \bibnamefont
  {Scully}}\ and\ \bibinfo {author} {\bibfnamefont {M.~S.}\ \bibnamefont
  {Zubairy}},\ }\href@noop {} {\emph {\bibinfo {title} {Quantum Optics}}}\
  (\bibinfo  {publisher} {AAPT},\ \bibinfo {year} {1999})\BibitemShut {NoStop}%
\bibitem [{\citenamefont {Ludlow}\ \emph {et~al.}(2015)\citenamefont {Ludlow},
  \citenamefont {Boyd}, \citenamefont {Ye}, \citenamefont {Peik},\ and\
  \citenamefont {Schmidt}}]{ludlow2015optical}%
  \BibitemOpen
  \bibfield  {author} {\bibinfo {author} {\bibfnamefont {A.~D.}\ \bibnamefont
  {Ludlow}}, \bibinfo {author} {\bibfnamefont {M.~M.}\ \bibnamefont {Boyd}},
  \bibinfo {author} {\bibfnamefont {J.}~\bibnamefont {Ye}}, \bibinfo {author}
  {\bibfnamefont {E.}~\bibnamefont {Peik}}, \ and\ \bibinfo {author}
  {\bibfnamefont {P.}~\bibnamefont {Schmidt}},\ }\href {\doibase
  10.1103/revmodphys.87.637} {\bibfield  {journal} {\bibinfo  {journal}
  {Reviews of Modern Physics}\ }\textbf {\bibinfo {volume} {87}},\ \bibinfo
  {pages} {637} (\bibinfo {year} {2015})}\BibitemShut {NoStop}%
\bibitem [{\citenamefont {Metcalf}\ and\ \citenamefont {Van~der
  Straten}(2007)}]{metcalf2007}%
  \BibitemOpen
  \bibfield  {author} {\bibinfo {author} {\bibfnamefont {H.~J.}\ \bibnamefont
  {Metcalf}}\ and\ \bibinfo {author} {\bibfnamefont {P.}~\bibnamefont {Van~der
  Straten}},\ }\href@noop {} {\emph {\bibinfo {title} {Laser Cooling and
  Trapping of Neutral Atoms}}}\ (\bibinfo  {publisher} {Wiley Online Library},\
  \bibinfo {year} {2007})\BibitemShut {NoStop}%
\bibitem [{\citenamefont {Su}\ \emph {et~al.}(2014)\citenamefont {Su},
  \citenamefont {Meng}, \citenamefont {Ji}, \citenamefont {Yuan}, \citenamefont
  {Zhao}, \citenamefont {Xiao},\ and\ \citenamefont {Jia}}]{Su2014}%
  \BibitemOpen
  \bibfield  {author} {\bibinfo {author} {\bibfnamefont {D.-Q.}\ \bibnamefont
  {Su}}, \bibinfo {author} {\bibfnamefont {T.-F.}\ \bibnamefont {Meng}},
  \bibinfo {author} {\bibfnamefont {Z.-H.}\ \bibnamefont {Ji}}, \bibinfo
  {author} {\bibfnamefont {J.-P.}\ \bibnamefont {Yuan}}, \bibinfo {author}
  {\bibfnamefont {Y.-T.}\ \bibnamefont {Zhao}}, \bibinfo {author}
  {\bibfnamefont {L.-T.}\ \bibnamefont {Xiao}}, \ and\ \bibinfo {author}
  {\bibfnamefont {S.-T.}\ \bibnamefont {Jia}},\ }\href {\doibase
  10.1364/ao.53.007011} {\bibfield  {journal} {\bibinfo  {journal} {Applied
  Optics}\ }\textbf {\bibinfo {volume} {53}},\ \bibinfo {pages} {7011}
  (\bibinfo {year} {2014})}\BibitemShut {NoStop}%
\bibitem [{\citenamefont {Zi}\ \emph {et~al.}(2017)\citenamefont {Zi},
  \citenamefont {Wu}, \citenamefont {Zhong}, \citenamefont {Parker},
  \citenamefont {Yu}, \citenamefont {Budker}, \citenamefont {Lu},\ and\
  \citenamefont {M\"{u}ller}}]{Zi2017}%
  \BibitemOpen
  \bibfield  {author} {\bibinfo {author} {\bibfnamefont {F.}~\bibnamefont
  {Zi}}, \bibinfo {author} {\bibfnamefont {X.}~\bibnamefont {Wu}}, \bibinfo
  {author} {\bibfnamefont {W.}~\bibnamefont {Zhong}}, \bibinfo {author}
  {\bibfnamefont {R.~H.}\ \bibnamefont {Parker}}, \bibinfo {author}
  {\bibfnamefont {C.}~\bibnamefont {Yu}}, \bibinfo {author} {\bibfnamefont
  {S.}~\bibnamefont {Budker}}, \bibinfo {author} {\bibfnamefont
  {X.}~\bibnamefont {Lu}}, \ and\ \bibinfo {author} {\bibfnamefont
  {H.}~\bibnamefont {M\"{u}ller}},\ }\href {\doibase 10.1364/ao.56.002649}
  {\bibfield  {journal} {\bibinfo  {journal} {Applied Optics}\ }\textbf
  {\bibinfo {volume} {56}},\ \bibinfo {pages} {2649} (\bibinfo {year}
  {2017})}\BibitemShut {NoStop}%
\bibitem [{\citenamefont {Harris}\ \emph {et~al.}(2006)\citenamefont {Harris},
  \citenamefont {Adams}, \citenamefont {Cornish}, \citenamefont {McLeod},
  \citenamefont {Tarleton},\ and\ \citenamefont {Hughes}}]{Harris2006}%
  \BibitemOpen
  \bibfield  {author} {\bibinfo {author} {\bibfnamefont {M.~L.}\ \bibnamefont
  {Harris}}, \bibinfo {author} {\bibfnamefont {C.~S.}\ \bibnamefont {Adams}},
  \bibinfo {author} {\bibfnamefont {S.~L.}\ \bibnamefont {Cornish}}, \bibinfo
  {author} {\bibfnamefont {I.~C.}\ \bibnamefont {McLeod}}, \bibinfo {author}
  {\bibfnamefont {E.}~\bibnamefont {Tarleton}}, \ and\ \bibinfo {author}
  {\bibfnamefont {I.~G.}\ \bibnamefont {Hughes}},\ }\href@noop {} {\bibfield
  {journal} {\bibinfo  {journal} {Physical Review A}\ }\textbf {\bibinfo
  {volume} {73}} (\bibinfo {year} {2006})}\BibitemShut {NoStop}%
\bibitem [{\citenamefont {Bjorklund}\ \emph {et~al.}(1983)\citenamefont
  {Bjorklund}, \citenamefont {Levenson}, \citenamefont {Lenth},\ and\
  \citenamefont {Ortiz}}]{Bjorklund1983}%
  \BibitemOpen
  \bibfield  {author} {\bibinfo {author} {\bibfnamefont {G.~C.}\ \bibnamefont
  {Bjorklund}}, \bibinfo {author} {\bibfnamefont {M.~D.}\ \bibnamefont
  {Levenson}}, \bibinfo {author} {\bibfnamefont {W.}~\bibnamefont {Lenth}}, \
  and\ \bibinfo {author} {\bibfnamefont {C.}~\bibnamefont {Ortiz}},\ }\href
  {\doibase 10.1007/bf00688820} {\bibfield  {journal} {\bibinfo  {journal}
  {Applied Physics B Photophysics and Laser Chemistry}\ }\textbf {\bibinfo
  {volume} {32}},\ \bibinfo {pages} {145} (\bibinfo {year} {1983})}\BibitemShut
  {NoStop}%
\bibitem [{\citenamefont {McCarron}, \citenamefont {King},\ and\ \citenamefont
  {Cornish}(2008)}]{McCarron2008}%
  \BibitemOpen
  \bibfield  {author} {\bibinfo {author} {\bibfnamefont {D.~J.}\ \bibnamefont
  {McCarron}}, \bibinfo {author} {\bibfnamefont {S.~A.}\ \bibnamefont {King}},
  \ and\ \bibinfo {author} {\bibfnamefont {S.~L.}\ \bibnamefont {Cornish}},\
  }\href {\doibase 10.1088/0957-0233/19/10/105601} {\bibfield  {journal}
  {\bibinfo  {journal} {Measurement Science and Technology}\ }\textbf {\bibinfo
  {volume} {19}},\ \bibinfo {pages} {105601} (\bibinfo {year}
  {2008})}\BibitemShut {NoStop}%
\bibitem [{\citenamefont {Li}\ \emph {et~al.}(2004)\citenamefont {Li},
  \citenamefont {Ido}, \citenamefont {Eichler},\ and\ \citenamefont
  {Katori}}]{Li2004}%
  \BibitemOpen
  \bibfield  {author} {\bibinfo {author} {\bibfnamefont {Y.}~\bibnamefont
  {Li}}, \bibinfo {author} {\bibfnamefont {T.}~\bibnamefont {Ido}}, \bibinfo
  {author} {\bibfnamefont {T.}~\bibnamefont {Eichler}}, \ and\ \bibinfo
  {author} {\bibfnamefont {H.}~\bibnamefont {Katori}},\ }\href {\doibase
  10.1007/s00340-004-1405-x} {\bibfield  {journal} {\bibinfo  {journal}
  {Applied Physics B}\ }\textbf {\bibinfo {volume} {78}},\ \bibinfo {pages}
  {315} (\bibinfo {year} {2004})}\BibitemShut {NoStop}%
\bibitem [{\citenamefont {Bridge}\ \emph {et~al.}(2009)\citenamefont {Bridge},
  \citenamefont {Millen}, \citenamefont {Adams},\ and\ \citenamefont
  {Jones}}]{Bridge2009}%
  \BibitemOpen
  \bibfield  {author} {\bibinfo {author} {\bibfnamefont {E.~M.}\ \bibnamefont
  {Bridge}}, \bibinfo {author} {\bibfnamefont {J.}~\bibnamefont {Millen}},
  \bibinfo {author} {\bibfnamefont {C.~S.}\ \bibnamefont {Adams}}, \ and\
  \bibinfo {author} {\bibfnamefont {M.~P.~A.}\ \bibnamefont {Jones}},\ }\href
  {\doibase 10.1063/1.3036980} {\bibfield  {journal} {\bibinfo  {journal}
  {Review of Scientific Instruments}\ }\textbf {\bibinfo {volume} {80}},\
  \bibinfo {pages} {013101} (\bibinfo {year} {2009})}\BibitemShut {NoStop}%
\bibitem [{\citenamefont {Norcia}\ and\ \citenamefont
  {Thompson}(2016)}]{Norcia2016}%
  \BibitemOpen
  \bibfield  {author} {\bibinfo {author} {\bibfnamefont {M.~A.}\ \bibnamefont
  {Norcia}}\ and\ \bibinfo {author} {\bibfnamefont {J.~K.}\ \bibnamefont
  {Thompson}},\ }\href {\doibase 10.1063/1.4942434} {\bibfield  {journal}
  {\bibinfo  {journal} {Review of Scientific Instruments}\ }\textbf {\bibinfo
  {volume} {87}},\ \bibinfo {pages} {023110} (\bibinfo {year}
  {2016})}\BibitemShut {NoStop}%
\bibitem [{\citenamefont {Lee}\ \emph {et~al.}(2014)\citenamefont {Lee},
  \citenamefont {Jarratt}, \citenamefont {Marciniak},\ and\ \citenamefont
  {Biercuk}}]{Lee2014}%
  \BibitemOpen
  \bibfield  {author} {\bibinfo {author} {\bibfnamefont {M.~W.}\ \bibnamefont
  {Lee}}, \bibinfo {author} {\bibfnamefont {M.~C.}\ \bibnamefont {Jarratt}},
  \bibinfo {author} {\bibfnamefont {C.}~\bibnamefont {Marciniak}}, \ and\
  \bibinfo {author} {\bibfnamefont {M.~J.}\ \bibnamefont {Biercuk}},\ }\href
  {\doibase 10.1364/oe.22.007210} {\bibfield  {journal} {\bibinfo  {journal}
  {Optics Express}\ }\textbf {\bibinfo {volume} {22}},\ \bibinfo {pages} {7210}
  (\bibinfo {year} {2014})}\BibitemShut {NoStop}%
\bibitem [{\citenamefont {Talvitie}, \citenamefont {Merimaa},\ and\
  \citenamefont {Ikonen}(1998)}]{Talvitie1998}%
  \BibitemOpen
  \bibfield  {author} {\bibinfo {author} {\bibfnamefont {H.}~\bibnamefont
  {Talvitie}}, \bibinfo {author} {\bibfnamefont {M.}~\bibnamefont {Merimaa}}, \
  and\ \bibinfo {author} {\bibfnamefont {E.}~\bibnamefont {Ikonen}},\ }\href
  {\doibase 10.1016/s0030-4018(98)00166-7} {\bibfield  {journal} {\bibinfo
  {journal} {Optics Communications}\ }\textbf {\bibinfo {volume} {152}},\
  \bibinfo {pages} {182} (\bibinfo {year} {1998})}\BibitemShut {NoStop}%
\bibitem [{\citenamefont {Vasilyev}\ \emph {et~al.}(2011)\citenamefont
  {Vasilyev}, \citenamefont {Nevsky}, \citenamefont {Ernsting}, \citenamefont
  {Hansen}, \citenamefont {Shen},\ and\ \citenamefont
  {Schiller}}]{Vasilyev2011}%
  \BibitemOpen
  \bibfield  {author} {\bibinfo {author} {\bibfnamefont {S.}~\bibnamefont
  {Vasilyev}}, \bibinfo {author} {\bibfnamefont {A.}~\bibnamefont {Nevsky}},
  \bibinfo {author} {\bibfnamefont {I.}~\bibnamefont {Ernsting}}, \bibinfo
  {author} {\bibfnamefont {M.}~\bibnamefont {Hansen}}, \bibinfo {author}
  {\bibfnamefont {J.}~\bibnamefont {Shen}}, \ and\ \bibinfo {author}
  {\bibfnamefont {S.}~\bibnamefont {Schiller}},\ }\href {\doibase
  10.1007/s00340-011-4435-1} {\bibfield  {journal} {\bibinfo  {journal}
  {Applied Physics B}\ }\textbf {\bibinfo {volume} {103}},\ \bibinfo {pages}
  {27} (\bibinfo {year} {2011})}\BibitemShut {NoStop}%
\bibitem [{\citenamefont {Tonyushkin}, \citenamefont {Light},\ and\
  \citenamefont {Rosa}(2007)}]{tonyushkin2007}%
  \BibitemOpen
  \bibfield  {author} {\bibinfo {author} {\bibfnamefont {A.~A.}\ \bibnamefont
  {Tonyushkin}}, \bibinfo {author} {\bibfnamefont {A.~D.}\ \bibnamefont
  {Light}}, \ and\ \bibinfo {author} {\bibfnamefont {M.~D.~D.}\ \bibnamefont
  {Rosa}},\ }\href {\doibase 10.1063/1.2818773} {\bibfield  {journal} {\bibinfo
   {journal} {Review of Scientific Instruments}\ }\textbf {\bibinfo {volume}
  {78}},\ \bibinfo {pages} {123103} (\bibinfo {year} {2007})}\BibitemShut
  {NoStop}%
\bibitem [{\citenamefont {Rossi}\ \emph {et~al.}(2002)\citenamefont {Rossi},
  \citenamefont {Biancalana}, \citenamefont {Mai},\ and\ \citenamefont
  {Tomassetti}}]{Rossi2002a}%
  \BibitemOpen
  \bibfield  {author} {\bibinfo {author} {\bibfnamefont {A.}~\bibnamefont
  {Rossi}}, \bibinfo {author} {\bibfnamefont {V.}~\bibnamefont {Biancalana}},
  \bibinfo {author} {\bibfnamefont {B.}~\bibnamefont {Mai}}, \ and\ \bibinfo
  {author} {\bibfnamefont {L.}~\bibnamefont {Tomassetti}},\ }\href {\doibase
  10.1063/1.1487895} {\bibfield  {journal} {\bibinfo  {journal} {Review of
  Scientific Instruments}\ }\textbf {\bibinfo {volume} {73}},\ \bibinfo {pages}
  {2544} (\bibinfo {year} {2002})}\BibitemShut {NoStop}%
\bibitem [{\citenamefont {Uetake}\ \emph {et~al.}(2009)\citenamefont {Uetake},
  \citenamefont {Matsubara}, \citenamefont {Ito}, \citenamefont {Hayasaka},\
  and\ \citenamefont {Hosokawa}}]{Uetake2009}%
  \BibitemOpen
  \bibfield  {author} {\bibinfo {author} {\bibfnamefont {S.}~\bibnamefont
  {Uetake}}, \bibinfo {author} {\bibfnamefont {K.}~\bibnamefont {Matsubara}},
  \bibinfo {author} {\bibfnamefont {H.}~\bibnamefont {Ito}}, \bibinfo {author}
  {\bibfnamefont {K.}~\bibnamefont {Hayasaka}}, \ and\ \bibinfo {author}
  {\bibfnamefont {M.}~\bibnamefont {Hosokawa}},\ }\href {\doibase
  10.1007/s00340-009-3619-4} {\bibfield  {journal} {\bibinfo  {journal}
  {Applied Physics B}\ }\textbf {\bibinfo {volume} {97}},\ \bibinfo {pages}
  {413} (\bibinfo {year} {2009})}\BibitemShut {NoStop}%
\bibitem [{\citenamefont {Black}(2001)}]{Black2001}%
  \BibitemOpen
  \bibfield  {author} {\bibinfo {author} {\bibfnamefont {E.~D.}\ \bibnamefont
  {Black}},\ }\href {\doibase 10.1119/1.1286663} {\bibfield  {journal}
  {\bibinfo  {journal} {American Journal of Physics}\ }\textbf {\bibinfo
  {volume} {69}},\ \bibinfo {pages} {79} (\bibinfo {year} {2001})}\BibitemShut
  {NoStop}%
\bibitem [{\citenamefont {Alnis}\ \emph {et~al.}(2008)\citenamefont {Alnis},
  \citenamefont {Matveev}, \citenamefont {Kolachevsky}, \citenamefont {Udem},\
  and\ \citenamefont {H{\"a}nsch}}]{alnis2008}%
  \BibitemOpen
  \bibfield  {author} {\bibinfo {author} {\bibfnamefont {J.}~\bibnamefont
  {Alnis}}, \bibinfo {author} {\bibfnamefont {A.}~\bibnamefont {Matveev}},
  \bibinfo {author} {\bibfnamefont {N.}~\bibnamefont {Kolachevsky}}, \bibinfo
  {author} {\bibfnamefont {T.}~\bibnamefont {Udem}}, \ and\ \bibinfo {author}
  {\bibfnamefont {T.}~\bibnamefont {H{\"a}nsch}},\ }\href {\doibase
  10.1103/physreva.77.053809} {\bibfield  {journal} {\bibinfo  {journal}
  {Physical Review A}\ }\textbf {\bibinfo {volume} {77}},\ \bibinfo {pages}
  {053809} (\bibinfo {year} {2008})}\BibitemShut {NoStop}%
\bibitem [{\citenamefont {Kobtsev}, \citenamefont {Kandrushin},\ and\
  \citenamefont {Potekhin}(2007)}]{kobtsev2007}%
  \BibitemOpen
  \bibfield  {author} {\bibinfo {author} {\bibfnamefont {S.}~\bibnamefont
  {Kobtsev}}, \bibinfo {author} {\bibfnamefont {S.}~\bibnamefont {Kandrushin}},
  \ and\ \bibinfo {author} {\bibfnamefont {A.}~\bibnamefont {Potekhin}},\
  }\href {\doibase 10.1364/ao.46.005840} {\bibfield  {journal} {\bibinfo
  {journal} {Applied Optics}\ }\textbf {\bibinfo {volume} {46}},\ \bibinfo
  {pages} {5840} (\bibinfo {year} {2007})}\BibitemShut {NoStop}%
\bibitem [{\citenamefont {Metzger}\ \emph {et~al.}(2017)\citenamefont
  {Metzger}, \citenamefont {Spesyvtsev}, \citenamefont {Bruce}, \citenamefont
  {Miller}, \citenamefont {Maker}, \citenamefont {Malcolm}, \citenamefont
  {Mazilu},\ and\ \citenamefont {Dholakia}}]{Metzger2017}%
  \BibitemOpen
  \bibfield  {author} {\bibinfo {author} {\bibfnamefont {N.~K.}\ \bibnamefont
  {Metzger}}, \bibinfo {author} {\bibfnamefont {R.}~\bibnamefont {Spesyvtsev}},
  \bibinfo {author} {\bibfnamefont {G.~D.}\ \bibnamefont {Bruce}}, \bibinfo
  {author} {\bibfnamefont {B.}~\bibnamefont {Miller}}, \bibinfo {author}
  {\bibfnamefont {G.~T.}\ \bibnamefont {Maker}}, \bibinfo {author}
  {\bibfnamefont {G.}~\bibnamefont {Malcolm}}, \bibinfo {author} {\bibfnamefont
  {M.}~\bibnamefont {Mazilu}}, \ and\ \bibinfo {author} {\bibfnamefont
  {K.}~\bibnamefont {Dholakia}},\ }\href {\doibase 10.1038/ncomms15610}
  {\bibfield  {journal} {\bibinfo  {journal} {Nature Communications}\ }\textbf
  {\bibinfo {volume} {8}},\ \bibinfo {pages} {15610} (\bibinfo {year}
  {2017})}\BibitemShut {NoStop}%
\bibitem [{\citenamefont {Saleh}\ \emph {et~al.}(2015)\citenamefont {Saleh},
  \citenamefont {Millo}, \citenamefont {Didier}, \citenamefont
  {Kersal{\'{e}}},\ and\ \citenamefont {Lacro{\^{u}}te}}]{saleh2015}%
  \BibitemOpen
  \bibfield  {author} {\bibinfo {author} {\bibfnamefont {K.}~\bibnamefont
  {Saleh}}, \bibinfo {author} {\bibfnamefont {J.}~\bibnamefont {Millo}},
  \bibinfo {author} {\bibfnamefont {A.}~\bibnamefont {Didier}}, \bibinfo
  {author} {\bibfnamefont {Y.}~\bibnamefont {Kersal{\'{e}}}}, \ and\ \bibinfo
  {author} {\bibfnamefont {C.}~\bibnamefont {Lacro{\^{u}}te}},\ }\href
  {\doibase 10.1364/ao.54.009446} {\bibfield  {journal} {\bibinfo  {journal}
  {Applied Optics}\ }\textbf {\bibinfo {volume} {54}},\ \bibinfo {pages} {9446}
  (\bibinfo {year} {2015})}\BibitemShut {NoStop}%
\bibitem [{\citenamefont {Sansonetti}\ and\ \citenamefont
  {Martin}(2005)}]{Sansonetti2005}%
  \BibitemOpen
  \bibfield  {author} {\bibinfo {author} {\bibfnamefont {J.~E.}\ \bibnamefont
  {Sansonetti}}\ and\ \bibinfo {author} {\bibfnamefont {W.~C.}\ \bibnamefont
  {Martin}},\ }\href {\doibase 10.1063/1.1800011} {\bibfield  {journal}
  {\bibinfo  {journal} {Journal of Physical and Chemical Reference Data}\
  }\textbf {\bibinfo {volume} {34}},\ \bibinfo {pages} {1559} (\bibinfo {year}
  {2005})}\BibitemShut {NoStop}%
\bibitem [{\citenamefont {Saakyan}\ \emph {et~al.}(2015)\citenamefont
  {Saakyan}, \citenamefont {Sautenkov}, \citenamefont {Vilshanskaya},
  \citenamefont {Vasiliev}, \citenamefont {Zelener},\ and\ \citenamefont
  {Zelener}}]{saakyan2015}%
  \BibitemOpen
  \bibfield  {author} {\bibinfo {author} {\bibfnamefont {S.~A.}\ \bibnamefont
  {Saakyan}}, \bibinfo {author} {\bibfnamefont {V.~A.}\ \bibnamefont
  {Sautenkov}}, \bibinfo {author} {\bibfnamefont {E.~V.}\ \bibnamefont
  {Vilshanskaya}}, \bibinfo {author} {\bibfnamefont {V.~V.}\ \bibnamefont
  {Vasiliev}}, \bibinfo {author} {\bibfnamefont {B.~B.}\ \bibnamefont
  {Zelener}}, \ and\ \bibinfo {author} {\bibfnamefont {B.~V.}\ \bibnamefont
  {Zelener}},\ }\href {\doibase 10.1070/qe2015v045n09abeh015708} {\bibfield
  {journal} {\bibinfo  {journal} {Quantum Electronics}\ }\textbf {\bibinfo
  {volume} {45}},\ \bibinfo {pages} {828} (\bibinfo {year} {2015})}\BibitemShut
  {NoStop}%
\bibitem [{\citenamefont {Barwood}\ \emph {et~al.}(2012)\citenamefont
  {Barwood}, \citenamefont {Gill}, \citenamefont {Huang},\ and\ \citenamefont
  {Klein}}]{Barwood2012}%
  \BibitemOpen
  \bibfield  {author} {\bibinfo {author} {\bibfnamefont {G.~P.}\ \bibnamefont
  {Barwood}}, \bibinfo {author} {\bibfnamefont {P.}~\bibnamefont {Gill}},
  \bibinfo {author} {\bibfnamefont {G.}~\bibnamefont {Huang}}, \ and\ \bibinfo
  {author} {\bibfnamefont {H.~A.}\ \bibnamefont {Klein}},\ }\href {\doibase
  10.1088/0957-0233/23/5/055201} {\bibfield  {journal} {\bibinfo  {journal}
  {Measurement Science and Technology}\ }\textbf {\bibinfo {volume} {23}},\
  \bibinfo {pages} {055201} (\bibinfo {year} {2012})}\BibitemShut {NoStop}%
\bibitem [{\citenamefont {Riley}(2008)}]{riley2008handbook}%
  \BibitemOpen
  \bibfield  {author} {\bibinfo {author} {\bibfnamefont {W.}~\bibnamefont
  {Riley}},\ }\href@noop {} {\emph {\bibinfo {title} {Handbook of Frequency
  Stability Analysis}}}\ (\bibinfo  {publisher} {NIST},\ \bibinfo {year}
  {2008})\BibitemShut {NoStop}%
\bibitem [{\citenamefont {Scholl}\ \emph {et~al.}(2004)\citenamefont {Scholl},
  \citenamefont {Rehse}, \citenamefont {Holt},\ and\ \citenamefont
  {Rosner}}]{Scholl2004}%
  \BibitemOpen
  \bibfield  {author} {\bibinfo {author} {\bibfnamefont {T.~J.}\ \bibnamefont
  {Scholl}}, \bibinfo {author} {\bibfnamefont {S.~J.}\ \bibnamefont {Rehse}},
  \bibinfo {author} {\bibfnamefont {R.~A.}\ \bibnamefont {Holt}}, \ and\
  \bibinfo {author} {\bibfnamefont {S.~D.}\ \bibnamefont {Rosner}},\ }\href
  {\doibase 10.1063/1.1791871} {\bibfield  {journal} {\bibinfo  {journal}
  {Review of Scientific Instruments}\ }\textbf {\bibinfo {volume} {75}},\
  \bibinfo {pages} {3318} (\bibinfo {year} {2004})}\BibitemShut {NoStop}%
\bibitem [{\citenamefont {Banerjee}\ \emph {et~al.}(2001)\citenamefont
  {Banerjee}, \citenamefont {Rapol}, \citenamefont {Wasan},\ and\ \citenamefont
  {Natarajan}}]{Banerjee2001}%
  \BibitemOpen
  \bibfield  {author} {\bibinfo {author} {\bibfnamefont {A.}~\bibnamefont
  {Banerjee}}, \bibinfo {author} {\bibfnamefont {U.~D.}\ \bibnamefont {Rapol}},
  \bibinfo {author} {\bibfnamefont {A.}~\bibnamefont {Wasan}}, \ and\ \bibinfo
  {author} {\bibfnamefont {V.}~\bibnamefont {Natarajan}},\ }\href {\doibase
  10.1063/1.1408279} {\bibfield  {journal} {\bibinfo  {journal} {Applied
  Physics Letters}\ }\textbf {\bibinfo {volume} {79}},\ \bibinfo {pages} {2139}
  (\bibinfo {year} {2001})}\BibitemShut {NoStop}%
\bibitem [{\citenamefont {Reiser}\ and\ \citenamefont
  {Lopert}(1988)}]{reiser1988}%
  \BibitemOpen
  \bibfield  {author} {\bibinfo {author} {\bibfnamefont {C.}~\bibnamefont
  {Reiser}}\ and\ \bibinfo {author} {\bibfnamefont {R.~B.}\ \bibnamefont
  {Lopert}},\ }\href {\doibase 10.1364/ao.27.003656} {\bibfield  {journal}
  {\bibinfo  {journal} {Applied Optics}\ }\textbf {\bibinfo {volume} {27}},\
  \bibinfo {pages} {3656} (\bibinfo {year} {1988})}\BibitemShut {NoStop}%
\bibitem [{\citenamefont {Gray}, \citenamefont {Smith},\ and\ \citenamefont
  {Dunning}(1986)}]{gray1986}%
  \BibitemOpen
  \bibfield  {author} {\bibinfo {author} {\bibfnamefont {D.~F.}\ \bibnamefont
  {Gray}}, \bibinfo {author} {\bibfnamefont {K.~A.}\ \bibnamefont {Smith}}, \
  and\ \bibinfo {author} {\bibfnamefont {F.~B.}\ \bibnamefont {Dunning}},\
  }\href {\doibase 10.1364/ao.25.001339} {\bibfield  {journal} {\bibinfo
  {journal} {Applied Optics}\ }\textbf {\bibinfo {volume} {25}},\ \bibinfo
  {pages} {1339} (\bibinfo {year} {1986})}\BibitemShut {NoStop}%
\bibitem [{\citenamefont {Gardner}(1985)}]{Gardner1985}%
  \BibitemOpen
  \bibfield  {author} {\bibinfo {author} {\bibfnamefont {J.~L.}\ \bibnamefont
  {Gardner}},\ }\href {\doibase 10.1364/ao.24.003570} {\bibfield  {journal}
  {\bibinfo  {journal} {Applied Optics}\ }\textbf {\bibinfo {volume} {24}},\
  \bibinfo {pages} {3570} (\bibinfo {year} {1985})}\BibitemShut {NoStop}%
\bibitem [{\citenamefont {Kajava}, \citenamefont {Lauranto},\ and\
  \citenamefont {Salomaa}(1993)}]{Kajava1993}%
  \BibitemOpen
  \bibfield  {author} {\bibinfo {author} {\bibfnamefont {T.~T.}\ \bibnamefont
  {Kajava}}, \bibinfo {author} {\bibfnamefont {H.~M.}\ \bibnamefont
  {Lauranto}}, \ and\ \bibinfo {author} {\bibfnamefont {R.~R.~E.}\ \bibnamefont
  {Salomaa}},\ }\href {\doibase 10.1364/josab.10.001980} {\bibfield  {journal}
  {\bibinfo  {journal} {Journal of the Optical Society of America B}\ }\textbf
  {\bibinfo {volume} {10}},\ \bibinfo {pages} {1980} (\bibinfo {year}
  {1993})}\BibitemShut {NoStop}%
\bibitem [{\citenamefont {Rogers}(1982)}]{Rogers1982}%
  \BibitemOpen
  \bibfield  {author} {\bibinfo {author} {\bibfnamefont {J.~R.}\ \bibnamefont
  {Rogers}},\ }\href {\doibase 10.1364/josa.72.000638} {\bibfield  {journal}
  {\bibinfo  {journal} {Journal of the Optical Society of America}\ }\textbf
  {\bibinfo {volume} {72}},\ \bibinfo {pages} {638} (\bibinfo {year}
  {1982})}\BibitemShut {NoStop}%
\bibitem [{\citenamefont {Kajava}, \citenamefont {Lauranto},\ and\
  \citenamefont {Friberg}(1994)}]{Kajava1994}%
  \BibitemOpen
  \bibfield  {author} {\bibinfo {author} {\bibfnamefont {T.~T.}\ \bibnamefont
  {Kajava}}, \bibinfo {author} {\bibfnamefont {H.~M.}\ \bibnamefont
  {Lauranto}}, \ and\ \bibinfo {author} {\bibfnamefont {A.~T.}\ \bibnamefont
  {Friberg}},\ }\href {\doibase 10.1364/josaa.11.002045} {\bibfield  {journal}
  {\bibinfo  {journal} {Journal of the Optical Society of America A}\ }\textbf
  {\bibinfo {volume} {11}},\ \bibinfo {pages} {2045} (\bibinfo {year}
  {1994})}\BibitemShut {NoStop}%
\bibitem [{\citenamefont {Nosske}\ \emph {et~al.}(2017)\citenamefont {Nosske},
  \citenamefont {Couturier}, \citenamefont {Hu}, \citenamefont {Tan},
  \citenamefont {Qiao}, \citenamefont {Blume}, \citenamefont {Jiang},
  \citenamefont {Chen},\ and\ \citenamefont {Weidem{\"u}ller}}]{nosske2017}%
  \BibitemOpen
  \bibfield  {author} {\bibinfo {author} {\bibfnamefont {I.}~\bibnamefont
  {Nosske}}, \bibinfo {author} {\bibfnamefont {L.}~\bibnamefont {Couturier}},
  \bibinfo {author} {\bibfnamefont {F.}~\bibnamefont {Hu}}, \bibinfo {author}
  {\bibfnamefont {C.}~\bibnamefont {Tan}}, \bibinfo {author} {\bibfnamefont
  {C.}~\bibnamefont {Qiao}}, \bibinfo {author} {\bibfnamefont {J.}~\bibnamefont
  {Blume}}, \bibinfo {author} {\bibfnamefont {Y.}~\bibnamefont {Jiang}},
  \bibinfo {author} {\bibfnamefont {P.}~\bibnamefont {Chen}}, \ and\ \bibinfo
  {author} {\bibfnamefont {M.}~\bibnamefont {Weidem{\"u}ller}},\ }\href@noop {}
  {\bibfield  {journal} {\bibinfo  {journal} {Physical Review A}\ }\textbf
  {\bibinfo {volume} {96}},\ \bibinfo {pages} {053415} (\bibinfo {year}
  {2017})}\BibitemShut {NoStop}%
\bibitem [{pri()}]{privcomm}%
  \BibitemOpen
  \href@noop {} {\bibinfo  {journal} {Private communication with HighFinesse}\
  }\BibitemShut {NoStop}%
\bibitem [{\citenamefont {Shimada}\ \emph {et~al.}(2013)\citenamefont
  {Shimada}, \citenamefont {Chida}, \citenamefont {Ohtsubo}, \citenamefont
  {Aoki}, \citenamefont {Takeuchi}, \citenamefont {Kuga},\ and\ \citenamefont
  {Torii}}]{shimada2013}%
  \BibitemOpen
\bibfield  {journal} {  }\bibfield  {author} {\bibinfo {author} {\bibfnamefont
  {Y.}~\bibnamefont {Shimada}}, \bibinfo {author} {\bibfnamefont
  {Y.}~\bibnamefont {Chida}}, \bibinfo {author} {\bibfnamefont
  {N.}~\bibnamefont {Ohtsubo}}, \bibinfo {author} {\bibfnamefont
  {T.}~\bibnamefont {Aoki}}, \bibinfo {author} {\bibfnamefont {M.}~\bibnamefont
  {Takeuchi}}, \bibinfo {author} {\bibfnamefont {T.}~\bibnamefont {Kuga}}, \
  and\ \bibinfo {author} {\bibfnamefont {Y.}~\bibnamefont {Torii}},\ }\href
  {\doibase 10.1063/1.4808246} {\bibfield  {journal} {\bibinfo  {journal}
  {Review of Scientific Instruments}\ }\textbf {\bibinfo {volume} {84}},\
  \bibinfo {pages} {063101} (\bibinfo {year} {2013})}\BibitemShut {NoStop}%
\bibitem [{\citenamefont {Sansonetti}\ and\ \citenamefont
  {Nave}(2010)}]{Sansonetti2010}%
  \BibitemOpen
  \bibfield  {author} {\bibinfo {author} {\bibfnamefont {J.~E.}\ \bibnamefont
  {Sansonetti}}\ and\ \bibinfo {author} {\bibfnamefont {G.}~\bibnamefont
  {Nave}},\ }\href {\doibase 10.1063/1.3449176} {\bibfield  {journal} {\bibinfo
   {journal} {Journal of Physical and Chemical Reference Data}\ }\textbf
  {\bibinfo {volume} {39}},\ \bibinfo {pages} {033103} (\bibinfo {year}
  {2010})}\BibitemShut {NoStop}%
\bibitem [{\citenamefont {Barlow}(1989)}]{Barlow1989}%
  \BibitemOpen
  \bibfield  {author} {\bibinfo {author} {\bibfnamefont {R.~J.}\ \bibnamefont
  {Barlow}},\ }\href@noop {} {\emph {\bibinfo {title} {Statistics: A Guide to
  the Use of Statistical Methods in the Physical Sciences}}},\ Vol.~\bibinfo
  {volume} {29}\ (\bibinfo  {publisher} {John Wiley \& Sons},\ \bibinfo {year}
  {1989})\BibitemShut {NoStop}%
\bibitem [{Hig()}]{HighFinesse}%
  \BibitemOpen
  \href@noop {} {\bibinfo  {journal}
  {http://www.highfinesse.com/en/wavelengthmeter/54/ws8-series}\ }\BibitemShut
  {NoStop}%
\end{thebibliography}%

\end{document}